\documentclass[pre,aps,twocolumn]{revtex4}
\usepackage{epsfig}
\usepackage{bm}
\newcommand{\B}[1]{{\bm{#1}}}

\usepackage[latin1]{inputenc}
\newcommand{\beq}{\begin{equation}}
\newcommand{\eeq}{\end{equation}}
\newcommand{\bea}{\begin{eqnarray}}
\newcommand{\eea}{\end{eqnarray}}
\begin{document}
\title{Stress field around arbitrarily shaped cracks in two-dimensional
elastic materials\\ Version of \today}
\author{Eran Bouchbinder$^1$, Joachim Mathiesen$^{1,2}$ and Itamar Procaccia$^{1}$}
\affiliation{$^1$Dept. of Chemical Physics, The Weizmann Institute
of Science, Rehovot 76100, Israel\\
 $^2$The Niels Bohr Institute, 17 Blegdamsvej, Copenhagen, Denmark}
\begin{abstract}
The calculation of the stress field around an arbitrarily shaped
crack in an infinite two-dimensional elastic medium is a
mathematically daunting problem. With the exception of few exactly
soluble crack shapes the available results are based on either
perturbative approaches or on combinations of analytic and 
numerical techniques. We present
here a general solution of this problem for any arbitrary crack.
Along the way we develop a method to compute the conformal map
from the exterior of a circle to the exterior of a line of arbitrary
shape, offering it as  a superior alternative to the
classical Schwartz-Cristoffel transformation. Our calculation
results in an accurate estimate of the full stress field and in
particular of the stress intensity factors $K_I$ and $K_{II}$ and
the $T$-stress which are essential in the theory of fracture.
\end{abstract}
\maketitle
\section{Introduction}
The existence of a crack in a stressed elastic medium does not
necessarily mean that a catastrophic failure is in sight;
otherwise many home owners would be in a constant state of panic.
Indeed, one of the major objectives of the theory of fracture
mechanics is to predict when the existence of a crack would
necessarily lead to the failure of materials. A crucial ingredient
of such a prediction is the determination of the state of
deformation of a given material in the presence of the said crack
under the effect of a given external loading. This calculation is
not available for arbitrarily shaped cracks even in two
dimensional elastic media. To explain the difficulty consider for
a moment an existing crack in an infinite two dimensional
medium. Under given load
conditions the displacement field $\B u(\B r,t)$ is giving rise to
the elastic strain tensor $\epsilon_{ij}$:
\begin{equation}
\epsilon_{ij}\equiv \frac{1}{2}\left(\frac{\partial u_i}{\partial x_j}
+\frac{\partial u_j}{\partial x_i}\right)
\ . \label{strain}
\end{equation}
In linear elasticity the stress tensor is related to the
strain tensor by \cite{86LL}:
\begin{equation}
\sigma_{ij} =\frac{E}{1+\nu} \left(\frac{\nu}{1-2\nu} \delta_{ij}\sum_k\epsilon_{kk} + \epsilon_{ij}
\right) \ . \label{stress}
\end{equation}
where $E$ and $\nu$ are the Young's modulus and the Poisson's ratio respectively. When the boundary
conditions at infinity include both opening and shear modes (with
respect to the straight crack) the stress field near the tip of
any crack has a universal form \cite{98F}, i.e.
\begin{equation}
\sigma_{ij}(r,\varphi)=\frac{K_I}{\sqrt{2\pi
r}}\Sigma^I_{ij}(\varphi)+ \frac{K_{II}}{\sqrt{2\pi
r}}\Sigma^{II}_{ij}(\varphi) \ . \label{uniform}
\end{equation}
Here $K_{I}$ and $K_{II}$ are the ``stress intensity factors" with
respect to the opening and shear modes, whereas
$\Sigma^I_{ij}(\varphi)$ and $\Sigma^{II}_{ij}(\varphi)$ are
universal angular functions common to all configurations and
loading conditions. Despite the simplicity of Eq. (\ref{uniform}) the calculation
of the stress intensity factors is not simple. Their numerical value depends on the
shape of the crack, its dynamical history and on the far field boundary conditions.  

For a {\em straight}
crack the stress intensity factors are known exactly \cite{53Mus,99B}, i.e.
\begin{equation}
K_{I} = \sigma^{\infty}_{yy}\sqrt{\pi a}\ , \quad
K_{II}= \sigma^{\infty}_{xy}\sqrt{\pi a}\ \ ,
\end{equation}
where $2a$ is the crack length and $\sigma^{\infty}_{ij}$ is the
uniform load at infinity. The criterion for failure is then the
famous Griffith-Irwin criterion \cite{57Irw},
\begin{equation}
\frac{K_I^2+K_{II}^2}{E} =\Gamma \ , \label{irwin}
\end{equation}
where $\Gamma$ is the fracture energy. The physical meaning of Eq. (\ref{irwin}) is that the
crack will initiate (and may cause failure) when the elastic energy
flowing from the stress field in the bulk to the tip region is at
least as large as the energy lost by lengthening the crack (bond
breaking or any other energy cost involved). In the case of the
straight crack failure will occur, for a given level of the
external load, at a critical crack length.

For cracks
of arbitrary shape we still expect both Eqs. (\ref{uniform}) and (\ref{irwin}) to remain valid.
The problem is then how to compute the stress
intensity factors $K_I$ and $K_{II}$ when the crack is {\em not}
straight. Indeed, a well known paper by Cotterell and Rice
\cite{80CR} offers such an analytic calculation for a {\em slightly curved}
crack in perturbation theory in the amount of deviation from the
straight crack. Many other works developed alternative hybrid numerical-analytical 
techniques like singular integral equations based either on dislocation distribution
or crack ``opening displacement functions" and superposition methods \cite{Methods}. A large body of
research is devoted to direct numerical techniques like the finite element method \cite{00DMDSB}.
In this paper we offer a
non-perturbative approach to the calculation of the {\em full}
stress field of an arbitrarily shaped crack based on a conformal mapping technique.
In particular, we
will be able to estimate accurately the stress intensity factors
and any other relevant quantities like the $T$-stress, a quantity
to be defined below (cf. Eq. (\ref{SIF})).
 In Sect. 2 we lay out the mathematical problem. In Sect. 3 we present a solution
based on the method of iterated conformal maps. This section includes also the 
general problem of finding the conformal map from the unit circle to a line
of an arbitrary shape. The section culminates in the calculation of the stress
intensity factors. In Sect. 4 we exemplify the method and compare it against
exactly soluble cases. Sect. 5 offers a short summary
and a discussion.


\section{Mathematical Formulation}

The theory of elastostatic fracture mechanics in brittle
continuous media is based on the equilibrium equations for an
isotropic elastic body \cite{86LL}
\begin{equation}
\frac{\partial \sigma_{ij}}{\partial x_j}= 0 . \label{equil}
\end{equation}
For in-plane modes of fractures, i.e. under plane-stress or
plane-strain conditions, one introduces the Airy stress potential
$U(x,y)$ such that
\begin{equation}
\sigma_{xx}=\frac{\partial^2 U}{\partial y^2} \ ; \sigma_{xy}=-
\frac{\partial^2 U}{\partial x\partial y} \ ;
\sigma_{yy}=\frac{\partial^2 U}{\partial x^2} \ . \label{sigU}
\end{equation}
Thus the set of Eq. (\ref{equil}), after simple manipulations,
translate to a Bi-Laplace equation for the Airy stress
potential $U(x,y)$ \cite{86LL}
\begin{equation}
\Delta \Delta U(x,y) =0 \ , \label{bilaplace}
\end{equation}
with the prescribed boundary conditions on the crack and on the
external boundaries of the material. At this point we choose to
focus on the case of uniform remote loadings and
traction-free crack boundaries. This choice, although not the most
general, is of great interest and will serve to elucidate our
method. Other solutions may be obtained by superposition. Thus, the boundary conditions at infinity, for the two
in-plane symmetry modes of fracture, are presented as
\begin{eqnarray}
&&\!\!\!\!\!\!\!\!\sigma_{xx}(\infty)=0\ ; \sigma_{yy}(\infty)=\sigma_\infty\ ;
 \sigma_{xy}(\infty)=0\quad \text{Mode I}\label{mode1}\\
&&\!\!\!\!\!\!\!\!\sigma_{xx}(\infty)=0\ ; \sigma_{yy}(\infty)=0\ ;
 \sigma_{xy}(\infty)=\sigma_\infty\quad \text{Mode II}\ . \nonumber\\ \label{mode2}
\end{eqnarray}
In Addition, the free boundary conditions on the crack are
expressed as
\begin{equation}
\sigma_{xn}(s)=\sigma_{yn}(s)=0  \ , \label{bcm12}
\end{equation}
where $s$ is the arc-length parametrization of the crack boundary and
the subscript $n$ denotes the out-ward normal direction.

The solution of the Bi-Laplace equation can be written in terms of
{\em two} analytic functions $\phi(z)$ and $\eta(z)$ as
\begin{equation}
U(x,y)= \Re [\bar z\varphi(z)+\eta(z)] \ . \label{Uphichi}
\end{equation}
In terms of these two analytic functions, using Eq. (\ref{sigU}),
the stress components are given by
\begin{eqnarray}
\sigma_{yy}(x,y)&=&\Re [2 \varphi'(z)+ \bar z\varphi''(z)+\eta''(z)]\nonumber\\
\sigma_{xx}(x,y)&=&\Re [2 \varphi'(z)-\bar z\varphi''(z)-\eta''(z)]\nonumber\\
\sigma_{xy}(x,y)&=&\Im [\bar z\varphi''(z)+\eta''(z)].
\label{components}
\end{eqnarray}
In order to compute the full stress field one should first
formulate the boundary conditions in terms of the analytic
functions $\varphi(z)$ and $\eta(z)$ and to remove the gauge
freedom in Eq. (\ref{Uphichi}). The boundary conditions Eq. (\ref{bcm12}), using Eq.
(\ref{sigU}), can be rewritten as \cite{53Mus}
\begin{equation}
\partial_t\left[\frac{\partial U}{\partial x}
+i\frac{\partial U}{\partial y}\right]=0\ . \label{bcU}
\end{equation}
Note that we do not have enough boundary conditions
to determine $U(x,y)$ uniquely. In fact we can allow in Eq. (\ref{Uphichi})
arbitrary transformations of the form
\begin{eqnarray}
\varphi &\rightarrow& \varphi +iCz+\gamma\nonumber\\
\psi &\rightarrow& \psi +\tilde\gamma \ , \quad\psi\equiv \eta'
\end{eqnarray}
where $C$ is a real constant and $\gamma$ and $\tilde\gamma$ are
complex constants. This provides five degrees of freedom in the
definition of the Airy potential. Two of these freedoms are
removed by choosing the gauge in Eq. (\ref{bcU}) according to
\begin{equation}
\frac{\partial U}{\partial x}
+i\frac{\partial U}{\partial y} = 0  \ , \quad \text{on the boundary}\ . \label{choice}
\end{equation}
It is important to stress that whatever the choice of the five
freedoms, the stress tensor is unaffected; see
\cite{53Mus} for an exhaustive discussion of this point. Computing Eq. (\ref{choice}) in
terms of Eq. (\ref{Uphichi}) we arrive at the boundary condition
\begin{equation}
\varphi(z)+z\overline{\varphi'(z)}+\overline{\psi(z)}=0 \ .
\label{bccrack}
\end{equation}
To proceed we represent $\varphi(z)$ and $\psi(z)$ in
Laurent expansion form:
\begin{eqnarray}
\varphi(z) &=&\varphi_1 z + \varphi_0
+\varphi_{-1}/z+\varphi_{-2}/z^2+\cdots \ , \nonumber\\
\psi(z) &=&\psi_1 z + \psi_0
+\psi_{-1}/z+\psi_{-2}/z^2+\cdots \ . \label{Laurentpp}
\end{eqnarray}
This form is in agreement with the boundary conditions at infinity that
disallow higher order terms in $z$.
One freedom is now used to choose $\varphi_1$ to be real and two more freedoms will allow
us later on to fix $\varphi_0$.
Then, using the boundary conditions (\ref{mode1}) and (\ref{mode2}), we find
\begin{eqnarray}
 \varphi_1&=&\frac{\sigma_{\infty}}{4}\ ;\quad
 \psi_1=\frac{\sigma_{\infty}}{2} \quad  \text{ Mode I}  \ ,\nonumber\\
 \varphi_1&=&0 \ ; \quad\quad \psi_1=i\sigma_\infty
\quad \text{ Mode II} \ . \label{p1p1}
\end{eqnarray}


\section{The Solution}
As said above, the direct determination of the stress tensor for a
given arbitrary shaped crack is difficult. To overcome the
difficulty we perform an intermediate step of determining the
conformal map from the exterior of the unit circle to the 
exterior of our given crack. Currently the best available approach for such
a task is the
Schwartz-Cristoffel transformation. Here we will present an alternative 
new approach for finding the wanted
conformal transformation, given in terms of a functional iteration of
fundamental conformal maps. The use of iterated
conformal maps was pioneered by Hastings and Levitov
\cite{98HL}; it was subsequently turned into a powerful tool for the study
of fractal and fracture growth patterns \cite{99DHOPSS,00DFHP,00DLP,01BJLMP,01BDLP,02BLP,02LP}. In
the next subsection we describe how, given a crack shape, to construct a conformal map from the
complex
$\omega$-plane to the physical $z$-plane  such that the conformal map $z
= \Phi(\omega )$ maps the exterior of  the unit circle in the 
$\omega$-plane to the exterior of the crack in the physical $z$-plane,
after $n$ directed growth steps. We draw the reader's attention to
the fact that this method is more general than its application in
this paper, and in fact we offer it as a superior method to the
Schwartz-Cristoffel transformation, with hitherto undetermined
potential applications in a variety of two-dimensional contexts.

\begin{figure}
\centering
\epsfig{width=.45\textwidth,file=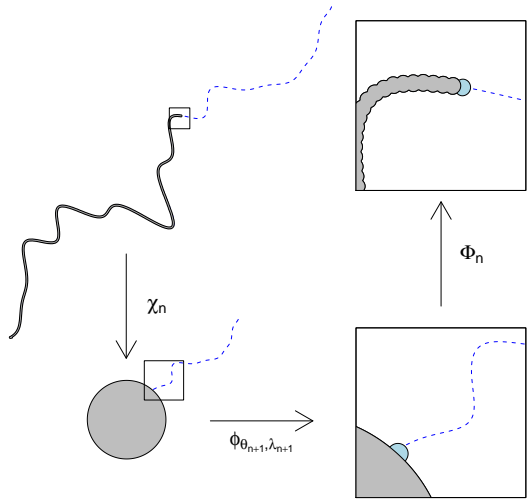}
\caption{Example of how to construct the conformal mapping along a line.}\label{method}
\end{figure}

\subsection{The conformal mapping}

The essential building block in the present application, as in all
the applications of the method of iterated conformal maps is the fundamental map
$\phi_{\lambda,\theta}$ that maps the exterior
circle onto the unit circle with a semi-circular bump of linear size
$\sqrt\lambda$ which is centered at the point $e^{i\theta}$. This  map reads
\cite{98HL}:
\begin{eqnarray}
\label{phi}
   &&\phi_{0,\lambda}(w) = \sqrt w \left\{ \frac{(1+
   \lambda)}{2w}(1+w)\right. \\
   &&\left.\times \left [ 1+w+w \left( 1+\frac{1}{w^2} -\frac{2}{w}
\frac{1-\lambda} {1+ \lambda} \right) ^{1/2} \right] -1 \right \} ^{1/2} \nonumber\\
   &&\phi_{\theta,\lambda} (w) = e^{i \theta} \phi_{0,\lambda}(e^{-i
   \theta}
   w) \,.
   \label{eq-f}
\end{eqnarray}
The inverse mapping
$\phi^{-1}_{\theta=0,\lambda}$ is of the form
\begin{equation}
  \label{eq:3}
  \phi^{-1}_{0,\lambda}= \frac{\lambda z-\sqrt{1+\lambda}(z^2-1)}{1-(1+\lambda)z^2}z \ .
\end{equation}
By composing this map with itself $n$ times with a judicious choice of
series $\{\theta_k\}_{k=1}^n$ and $\{\lambda_k\}_{n=1}^{n}$ we will construct
$\Phi^{(n)}(\omega)$ that will map the exterior of the circle to the exterior of an arbitrary
simply connected shape. To understand how to choose the two series $\{\theta_k\}_{k=1}^n$ and
$\{\lambda_k\}_{n=1}^{n}$ consider Fig.
\ref{method}, and define the inverse map $\omega=\chi^{(n)}(z)$.  Assume now that we already
have $\Phi^{(n-1)}(\omega)$ and therefore also its analytic inverse $\chi^{(n-1)}(z)$ after
$n-1$ growth steps, and we want to perform the next iteration. To construct
$\Phi^{(n)}(\omega)$ we advance our mapping in the direction of a point
$\tilde z$ in the
$z$-plane by adding a bump in the direction of $\tilde w=\chi^{(n-1)}(\tilde z)$ in the $w$-plane.
The map $\Phi^{(n)}(\omega)$ is obtained as follows:
\begin{equation}
\label{conformal}
\Phi^{(n)}(\omega) = \Phi^{(n-1)}(\phi_{\theta_n,\lambda_n}(\omega )) \ . \label{iter}
\end{equation}
The value of $\theta_n$ is determined by
\begin{equation}
  \label{eq:1}
  \theta_n=\arg [\chi^{(n-1)}(\tilde z)]
\end{equation}
The magnitude of the bump $\lambda_n$ is determined by requiring fixed size bumps
in the $z$-plane. This means that
\begin{equation}
  \label{eq:4}
  \lambda_n = \frac{\lambda_0}{|{\Phi^{(n-1)}}' (e^{i \theta_{n}})|^2}.
\end{equation}
We note here that it is not necessary in principle to have fixed
size bumps in the physical domain. In fact, adaptive size bumps
could lead to improvements in the precision and performance of our
scheme. We consider here the fixed size scheme for the sake of
simplicity, and we will show that the accuracy obtained is
sufficient for our purposes. Iterating the scheme
described above we end up with a conformal map that is written in
terms of an iteration over the fundamental maps (\ref{phi}):
\begin{equation}
  \label{eq:4}
  \Phi^{(n)}(w)=\phi_{\theta_{1},\lambda_{1}}\circ\ldots\circ\phi_{\theta_n,\lambda_n}(w) \ .
\end{equation}
For the sake of newcomers to the art of iterated conformal maps we stress that 
this iterative structure is abnormal, in the sense that the order 
of iterates in inverted with respect
to standard dynamical systems. On the other hand
the inverse mapping follows a standard iterative scheme
\begin{equation}
  \label{eq:4}
\chi^{(n)}(z)=\phi^{-1}_{\theta_{n},\lambda_{n}}\circ\ldots\circ\phi^{-1}_{\theta_1,\lambda_1}(z) \ .
\end{equation}

The algorithm is then described as follows; first we divide the
curve into segments separated by points $\{z_i\}$. The
spatial extent of each segment is taken to be approximately
$\sqrt{\lambda_0}$, in order to match the size of the bumps in the
$z$-plane. Without loss of generality we can take one of these
points to be at the center of coordinates and to be our starting
point. From the starting point we now advance along the shape by
mapping the next point $z_i$ on the curve according to the scheme
described above.


\subsection{Solution in terms of conformal mappings}

The conformal map $\Phi^{(n)}(\omega)$ is constructed in $n$ iterative steps. For
the discussion below we do not need the $n$ superscript and will denote simply
$\Phi(\omega ) \equiv {\Phi}^{(n)}(\omega )$. This map is
univalent \cite{99DHOPSS}, having the Laurent expansion form
\begin{equation}
\Phi(\omega ) = F_1\omega + F_0
+F_{-1}/\omega+F_{-2}/\omega^2+\cdots \ . \label{Laurent}
\end{equation}
Any position $z$ in
the physical domain is mapped by  $\chi(z)  \equiv {\Phi}^{-1}(z)$ onto a
position $\omega$ in the mathematical domain. This transformation
does not immediately provide the solution as the Bi-Laplacian
operator, in contrast to the Laplacian operator, is not
conformally invariant. Nevertheless, the conformal mapping method
can be extended to non-Laplacian problems. We begin by writing our
unknown functions $\varphi(z)$ and $\psi(z)$ in terms of the
conformal map
\begin{equation}
\varphi(z)\equiv \tilde \varphi\left(\chi(z)\right) \ , \quad
\psi(z)\equiv \tilde \psi\left(\chi(z)\right) \ . \label{ppp}
\end{equation}
Using the Laurent form of the conformal map, Eq. (\ref{Laurent}), the linear
term as $\omega\to \infty$ is determined by Eqs. (\ref{ppp}). We
therefore write
\begin{eqnarray}
\tilde \varphi(\omega) &=& \varphi_1F_1 \omega
+ \tilde\varphi_{-1}/\omega+\tilde\varphi_{-2}/\omega^2 +\dots\ , \nonumber\\
\tilde \psi(\omega) &=& \psi_1F_1 \omega + \tilde \psi_0
+\tilde\psi_{-1}/\omega +\tilde\psi_{-2}/\omega^2 +\dots\ ,
\label{expom}
\end{eqnarray}
where we used the last two freedoms to choose $\varphi_0=
-F_0\varphi_0$ such that $\tilde\varphi_0=0$. The boundary
condition (\ref{bccrack}) is now read for the unit circle in the
$\omega$-plane. Denoting $\epsilon\equiv e^{i\theta}$ and
\begin{equation}
u(\epsilon)\equiv \sum_{n=1}^\infty \tilde\varphi_{-n}/\epsilon^n \ ,
\quad v(\epsilon)\equiv  \sum_{n=0}^\infty \tilde\psi_{-n}/\epsilon^n \ , \label{uvdef}
\end{equation}
we write
\begin{equation}
u(\epsilon) +
\frac{\Phi(\epsilon)}{\overline{\Phi^{'}(\epsilon)}}
\overline{u'(\epsilon)}+\overline{v(\epsilon)}=f(\epsilon) \ .
\label{eq.fund}
\end{equation}
The function $f(\epsilon)$ is a known
function that contains all the coefficients that were determined
so far:
\begin{equation}
f(\epsilon) =-\varphi_1 F_1\epsilon - \frac{\Phi(\epsilon)}
{\overline{\Phi^{'}(\epsilon)}}\varphi_1 F_1
-\frac{\overline{\psi_1} F_1}{\epsilon} \label{ff} \ .
\end{equation}
To solve the problem we need to compute the coefficients $\tilde\varphi_n$ and $\tilde\psi_n$.
To this aim we first write \cite{02BLP}
\begin{equation}
\frac{\Phi(\epsilon)}{\overline{\Phi^{'}(\epsilon)}}=
\sum_{-\infty}^{\infty}b_i\epsilon^i. \label{expmap}
\end{equation}
The function $f(\epsilon)$ has also an expansion of the form
\begin{equation}
f(\epsilon)= \sum_{-\infty}^{\infty}f_{i}\epsilon^i \ . \label{fepsilon}
\end{equation}
In the discussion below we assume that the coefficients $b_i$ and
$f_i$ are known. In order to compute these coefficients we need to
Fourier transform the function $\Phi(\epsilon)/
\overline{\Phi^{'}(\epsilon)}$. This is the most expensive step
in our solution. One needs to carefully evaluate the Fourier
integrals between the branch cuts. Using the last two equations
together with Eqs. (\ref{uvdef}) and (\ref{eq.fund}) we obtain
\begin{eqnarray}
\tilde \varphi_{-m}&-&\sum_{k=1}^\infty k~b_{-m-k-1} \tilde\varphi^*_{-k} =f_{-m}
\ , \quad m=1,2\cdots \ , \label{power_sol1}\\
\tilde \psi^*_{-m}&-&\sum_{k=1}^\infty k~b_{m-k-1} \tilde\varphi^*_{-k} =f_{m}
\ . \quad m=0,1,2\cdots \ \label{power_sol2}
\end{eqnarray}
These sets of linear equations are well posed.  The coefficients
$\tilde \varphi_{-m}$ can be calculated from equation
(\ref{power_sol1}) alone,  and then they can be used to determine
the coefficients $\tilde \psi_{-m}$. This is in fact a proof that
Eq. (\ref{eq.fund}) determines the functions $u(\epsilon)$ and $v(\epsilon)$ together.
This fact had been proven with some generality in \cite{53Mus}.

The calculation of the Laurent expansion form of $\tilde
\varphi(\omega)$ and $\tilde \psi(\omega)$ provides the solution of
the problem in the $\omega$-plane. Still, one should express the
derivatives of $\varphi(z)$ and $\eta(z)$ in terms of $\tilde
\varphi(\omega)$ and $\tilde \psi(\omega)$ and the inverse map
$\chi(z)$ to obtain the solution in the physical $z$-plane. This
is straight forward and yields
\begin{eqnarray}
\varphi'(z)&=&\tilde\varphi'[\chi(z)] ~\chi'(z)\nonumber\\
\varphi''(z)&=&\tilde\varphi''[\chi(z)] ~[\chi'(z)]^2+
 \tilde\varphi'[\chi(z)] ~\chi''(z)\nonumber\\
\eta''(z)&=& \psi'(z)=\tilde\psi'[\chi(z)] ~\chi'(z).
\label{relations}
\end{eqnarray}
Upon substituting these relations into Eq. (\ref{components}) one
can calculate the {\em full} stress field for an
arbitrarily shaped crack. The expression of the stress field in
terms of the inverse conformal mapping is known for
quite a long time although it is very limited as the conformal
mapping and its inverse is rarely at hand. The central step of progress in
this paper is the conjunction of the novel functional iterative scheme
for obtaining the inverse conformal mapping
with the known result that expresses the stress field in
terms of this inverse mapping.


\subsection{The Stress Intensity Factors and the T-stress}

Linear elasticity fracture mechanics, under
small scale yielding, has no intrinsic length scale. Accordingly the stress
intensity factors are the most important quantities that
characterize the universal near-tip fields. At this point we explain
how to calculate the stress intensity factors from our solution. 

In principle, the calculation follows directly from the solution in terms of the
conformal map as described above. Previous authors derived the following expression
for the complex combination (of the real) stress intensity factors \cite{69And}:
\begin{equation}
K_I-iK_{II}= 2\sqrt{\frac{\pi}{e^{i\delta_j}\Phi''(\omega_j)}}~\tilde\varphi'(\omega_j),
\label{comlexSIF3}
\end{equation}
where $\omega_j$ is the position of the tip in the $\omega$-plane, and $\delta_j$ is the
argument of the tip position $z_j$ in the physical plane.
This result, although exact, cannot be used to obtain accurate estimates of the stress
intensity factors in our method. Since we construct our crack from a succession of little bumps,
our crack tip is not infinitely sharp. Therefore we cannot base our calculation of
the stress intensity factors on the precise coordinates of the tip. Rather, we need
to exploit our knowledge of the stress field for a substantial region around the tip.
We thus consider the stress field in the region of the tip, and write the components along the
tangent to the crack at the tip
\cite{98F}
\begin{eqnarray}
\sigma_{\varphi\varphi}(r,0)&=&\frac{K_I}{\sqrt{2\pi r}} + b_{\varphi\varphi}\sqrt{r} \nonumber\\
\sigma_{r \varphi}(r,0)&=&\frac{K_{II}}{\sqrt{2\pi r}} + b_{r \varphi}\sqrt{r} \nonumber\\
\sigma_{rr}(r,0)&=&\frac{K_{I}}{\sqrt{2\pi r}} + T +
b_{rr}\sqrt{r} \ . \label{SIF}
\end{eqnarray}
We have added terms of $O(\sqrt{r})$ to the leading terms, and in addition
we took into account the $T$-stress contribution to the purely radial component $\sigma_{rr}$.
It is crucial to note that there are no
other $O(1)$ terms in the first two lines of Eq. (\ref{SIF}). We did not need to consider 
explicitly any higher $O(r^{3/2})$ terms. In order to
extract the stress intensity factors from the full field
distributions we fit our solution near the tip to the form given
by Eq. (\ref{SIF}). We thus obtain not only the stress intensity factors but also
the coefficient of the subleading terms. We note that the so-called $T$-stress has an
important role in fracture theory (cf. \cite{80CR, 99B, 98F}). 

\section{Demonstrations of the Method}

In this section we demonstrate our method for two (of
the very rare) exactly soluble geometries. There are many numerically solved
problems in the literature that one can use for comparison, but our aim here is to
exemplify the essentials of our approach.
We will see that the method works very well and we propose that it can be used
for arbitrary crack shapes as well. 

\subsection{The Straight Crack}

The problem of a crack of length $2a$ in an infinite domain,
subjected to a remote uniaxial load $\sigma_{yy}=\sigma^{\infty}$
and traction-free crack faces is considered as the canonical
problem in the theory of linear elasticity fracture mechanics. The
resulting stress field is known, and is given by \cite{99B}:
\begin{eqnarray}
\sigma_{yy}(x,y)&=&\sigma^{\infty} \Re \left[\frac{z}{\sqrt{z^2-a^2}} +
\frac{i y a^2}{{(z^2-a^2)^{3/2}}}\right]\nonumber\\
\sigma_{xx}(x,y)&=&\sigma^{\infty} \Re \left[\frac{z}{\sqrt{z^2-a^2}} -
\frac{i y a^2}{{(z^2-a^2)^{3/2}}}\right]-\sigma^{\infty}\nonumber\\
\sigma_{xy}(x,y)&=&-\sigma^{\infty} \Re \left[
\frac{y a^2}{{(z^2-a^2)^{3/2}}}\right],
\label{StraightCrack}
\end{eqnarray}
where $z=x+iy$ and the crack is represented by a branch cut along
$-a<x<a, ~y=0$.

\begin{figure}
\centering
\epsfig{width=.45\textwidth,file=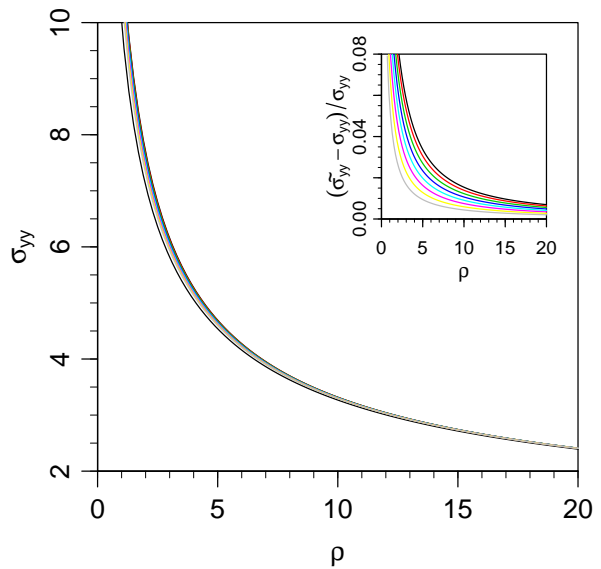}
\caption{The stress tensor component $\sigma_{yy}$
  for a straight crack with $a=200$. The component is evaluated for
  $y=0$ and a distance $\rho$ away from the tip at $x=200$. The numerical estimates $\tilde
  \sigma_{yy}$ approaches the analytical result (the left most line) as we decrease the
linear size of the bumps $\sqrt\lambda_0=1.0,0.9,0.8,\dots,0.3$. The inset
shows the relative error.}\label{line}
\end{figure}
We applied our general solution to the straight crack problem. We
constructed the conformal mapping using our functional iteration scheme
although the exact conformal mapping is known to be
\begin{equation}
\Phi(\omega)=\frac{a}{2}\left(\omega + \frac{1}{\omega}\right).
\end{equation}
Fig. \ref{line} compares our calculation of $\sigma_{yy}$ along the $x$-axis
with the exact result. The deviation near the tip of the crack is
expected as in our solution the crack tip is not infinitely
sharp but has a finite radius of curvature controlled by the
parameter $\lambda$ in Eq. (\ref{phi}). Decreasing the value of
$\lambda$, we can obtain more accurate results. Note
that real crack-tips {\em do} have a finite radius of curvature, so that our
method can be even more appropriate; the idealization of
representing cracks by mathematical branch cuts is no longer a
necessary simplification.

\begin{figure}
\centering
\epsfig{width=.45\textwidth,file=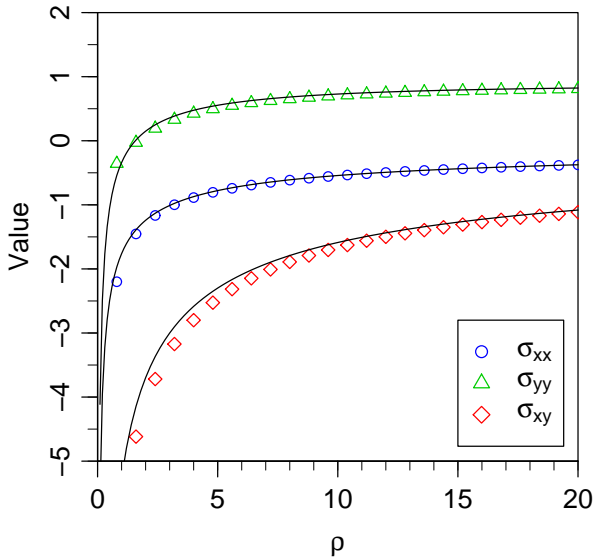}
\caption{The stress tensor components along the
  tangent to the crack tip. The crack is a semi circular arc of radius 200.  $\rho$
is the distance away from the tip. The points
  show the numerical values computed using bumps of linear size
  $\sqrt\lambda_0=0.6$ and the lines are the corresponding analytical values.}\label{circarc}
\end{figure}

\subsection{The Circular Arc Crack}

The analytic methods developed by Muskhelishvili \cite{53Mus}
lead to the solution of various problems concerned with circular
regions and infinite regions cut along circular arcs. Here we
consider a crack in the shape of a circular arc that extends from
$z=e^{i\theta}$ to $z=e^{-i\theta}$. This crack is subjected to a remote
uniaxial load $\sigma_{xx}=\sigma^{\infty}$ parallel to the $x$-axis
with traction-free crack boundaries. The stress tensor components are
calculated from Eq. (\ref{components}) with \cite{53Mus}
\begin{eqnarray}
\varphi'(z)&=&\frac{\sigma_{\infty}}{2~
G(z)}\left[C(\theta)(z-\cos(\theta))
+\frac{\cos(\theta)}{2z}-\frac{1}{2z^2}\right]\nonumber\\
&+&\sigma_{\infty}\left[\frac{1}{4}-\frac{C(\theta)}{2}-\frac{1}{4z^2}\right]\nonumber\\
\Omega(z)&=&\frac{\sigma_{\infty}}{2~
G(z)}\left[C(\theta)(z-\cos(\theta))
+\frac{\cos(\theta)}{2z}-\frac{1}{2z^2}\right]\nonumber\\
&-&\sigma_{\infty}\left[\frac{1}{4}-\frac{C(\theta)}{2}-\frac{1}{4z^2}\right]\nonumber\\
\psi'(z)&=&\frac{\varphi'(z)}{z^2}-\frac{\bar{\Omega}(1/z)}{z^2}-\frac{\varphi''(z)}{z}
\ . \label{CircArc}
\end{eqnarray}
Here
\begin{eqnarray}
C(\theta)&=&\frac{1-\sin^2{(\theta/2)}\cos^2{(\theta/2)}}
{2[1+\sin^2{(\theta/2)}]}\nonumber\\
G(z)&=&\sqrt{z^2-2z\cos(\theta)+1},
\end{eqnarray}
where $G(z)$ is defined such that $z^{-1}G(z)\rightarrow 1$ as
$z\rightarrow \infty$, which leads to $G(0)=-1$. Note that these
results are given for an arc of unit radius and therefore the solution
for any other circular arc can be obtained by a suitable
rescaling transformation.

We applied our method for this
configuration. The results for all the three components of the
stress tensor field for $\theta=\pi /2$ along the continuation of
the tip parallel to the $x$-axis are presented in Fig. \ref{circarc}
and compared with the exact results.

\begin{figure}
\centering
\epsfig{width=.45\textwidth,file=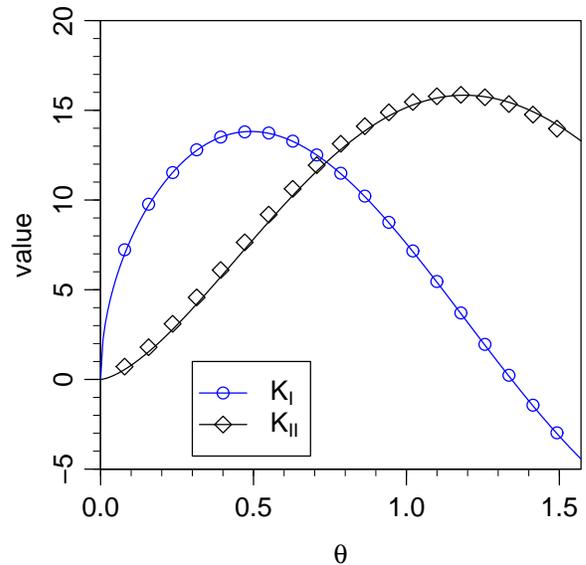}
\caption{Stress intensity factors for circular arcs of angles $2\theta$. The points are the
numerical values computed from fitting of the stress field along the tangent to the crack
tip. The fitting function used is of the form Eq. (\ref{SIF}). In this calculation
$\sqrt{\lambda_0}= 0.4$, and the fitting window is $5<\rho<30$.}\label{circarcSIF}
\end{figure}

Finally, we used the method to extract the stress intensity factors. 
Fig. \ref{circarcSIF} shows the mode I
and mode II stress intensity factors as a function of the half arc
angle $\theta$ and compares them with the exact analytic result
derived form the full field solution \cite{99B}.

\section{Summary and Conclusions}

In summary, we have demonstrated that the method of iterated conformal maps
can be used to construct the conformal map from the exterior of the unit circle
to the exterior of an arbitrary crack. Next one can address the calculation of the
stress field around such a crack with given loads at infinity. Having the said
conformal map at hand simplifies enormously the calculation of the full stress field,
allowing an accurate estimate of the stress intensity factors or of the 
sub-leading terms like the $T$-stress. The method was demonstrated by comparison with
exactly soluble examples. The quality of the comparison leads to the conclusion that the issue of
potential failure of a material given a crack and boundary conditions can be efficiently dealt
with. Future work will employ the present method to describe the dynamics of slow 
fracture where quasi-static methods are adequate.

\acknowledgments

This work had been supported in part by European Commission under a TMR grant.


\begin{thebibliography}{99}

\bibitem{86LL}
L.D. Landau and E.M. Lifshitz, {\em Theory of Elasticity}, 3rd ed.
(Pergamon, London, 1986).

\bibitem{98F}
L. B. Freund, {\em Dynamic Fracture Mechanics}, (Cambridge, 1998)

\bibitem{53Mus}
N. I. Muskhelishvili, {\em Some Basic Problems of the Mathematical
Theory of Elasticity}, (Noordhoff, 1953).

\bibitem{99B}
K. B.  Broberg, {\em Cracks and Fracture}, (Academic Press, 1999).

\bibitem{57Irw}
G. R. Irwin, J. App. Mech. {\bf 24}, 361 (1957).


\bibitem{80CR}
B. Cotterell and J. R. Rice, Int. J. Fract. {\bf16},   155 (1980).

\bibitem{Methods}
The number of publications on this problem is enormous. See for example, 
Y. Z. Chen, Eng. Fract. Mech. {\bf51},  97 (1995); Theor. Appl. Fract. Mech. {\bf31},  223
(1999); J. K. Burton and S. L. Phoenix, Int. J. Fract. {\bf102},  99 (2000).

\bibitem{00DMDSB}
See for example C. Daux, N. Mo\"es, John Dolbow, N. Sukumar and T. Belytschko,
Int. J. Numer. Meth. Engng, {\bf 48}, 1741 (2000).


\bibitem{98HL} M.B. Hastings and L.S. Levitov, Physica D {\bf 116},
244 (1998).

 \bibitem{99DHOPSS} B. Davidovitch, H.G.E. Hentschel, Z. Olami,
   I.Procaccia,
   L.M. Sander, and E. Somfai,
   Phys. Rev. E, {\bf 59} 1368 (1999).

\bibitem{00DFHP}
   B. Davidovitch, M.J. Feigenbaum, H.G.E. Hentschel and I. Procaccia,
   Phys. Rev. E {\bf 62}, 1706 (2000).


\bibitem{00DLP}
B. Davidovitch, A. Levermann, I. Procaccia, Phys. Rev. E, {\bf 62} R5919.

\bibitem{01BJLMP}
B. Davidovitch, M. H. Jensen, A. Levermann, J. Mathiesen and I.
Procaccia, Phys. Rev. Lett. {\bf 87}, 164101 (2001). 

  
 \bibitem{01BDLP}
F. Barra, B. Davidovitch, A. Levermann and I. Procaccia,
 Phys. Rev. Lett. {\bf 87}, p. 134501 (2001).

\bibitem{02BLP}
F. Barra, A. Levermann and I. Procaccia,  Phys. Rev. E., {\bf 66}, 066122 (2002).


\bibitem{02LP} 
A. Levermann and I. Procaccia, Phys. Rev. Lett.,  {\bf 89}, 234501 (2002).

\bibitem{69And}
H. Andersson, J. Mech. Phys. Solids {\bf 17}, 417 (1969).


\end{thebibliography}
\end{document}